\documentclass[a4paper]{jpconf}
\usepackage{graphicx}
\begin{document}
\title{Target, magnetic horn and safety studies for the CERN to Fr\'ejus Super Beam}

\author{  E Baussan, M Dracos, G Gaudiot, B Lepers, F Osswald, \\P Poussot, N Vassilopoulos, J Wurtz and V Zeter\\on behalf of EUROnu WP2 group }

\address{IPHC, Universit\'e de Strasbourg, CNRS/IN2P3 F-67037, Strasbourg, France}

\ead{nikolaos.vassilopoulos@iphc.cnrs.fr corresponding author}

\begin{abstract}
In the framework of the EUROnu design study, a new design for the CERN to Fr\'ejus neutrino beam based on
the SPL is under development by the WP2 group. 
The main challenge of this project lies with the design of a multi-MW neutrino beam facility. 
The horn and the decay tunnel parameters have been optimized  to maximize any potential discovery. The target design, thermo-mechanical analysis, and power supply design of the horn system as well as any safety issues are
being studied to meet the MW power requirements for the proton-beam. { \it (Contribution to NUFACT 11, XIIIth International Workshop on Neutrino Factories, Super Beams and Beta beams, 1-6 August 2011, CERN and University of Geneva (Submitted to IOP conference series))}
\end{abstract}


\section{Introduction \label{intro} }

The summary of the recent target and horn studies for the CERN to Fr\'ejus neutrino beam is presented in this paper. The main design and the physics reach of the Super Beam project  are described in \cite{physicsopt}. The optimization procedure for the horn shape and layout-geometry  to achieve optimum physics, the study of a target able to withstand a multi-MW proton-beam power, multi-physics simulation to investigate heat transfer, cooling and mechanical stress for the horn  
and safety aspects are discussed here. 

\section{The proton-beam and horn/target station \label{protonbeam} }

\begin{figure}[h]
\begin{minipage}{14pc}
\includegraphics[width=14pc]{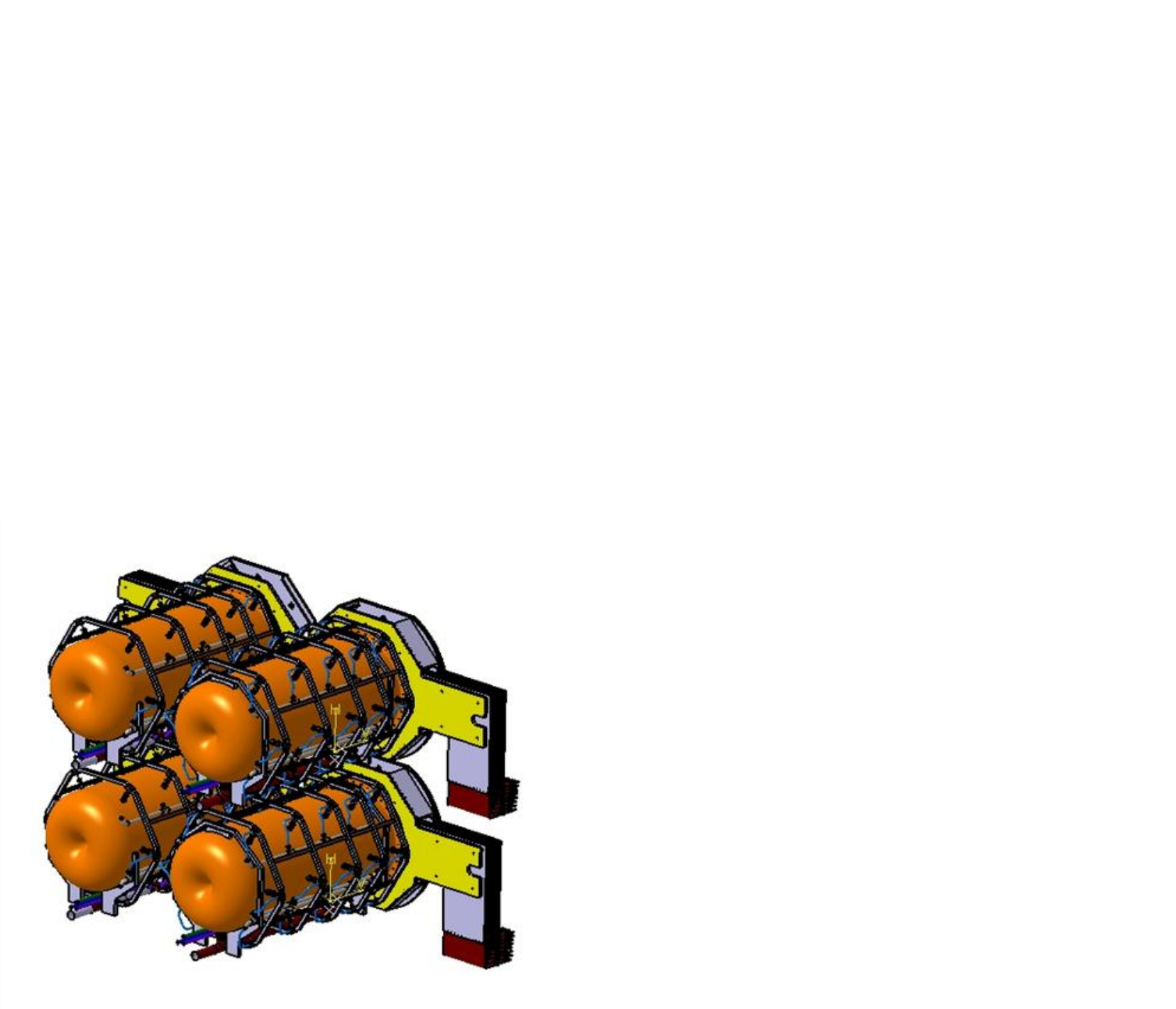}
\caption{\label{fig: 4horn} Four-horn assembly.}
\end{minipage}\hspace{2pc}%
\begin{minipage}{22pc}
\includegraphics[width=22pc]{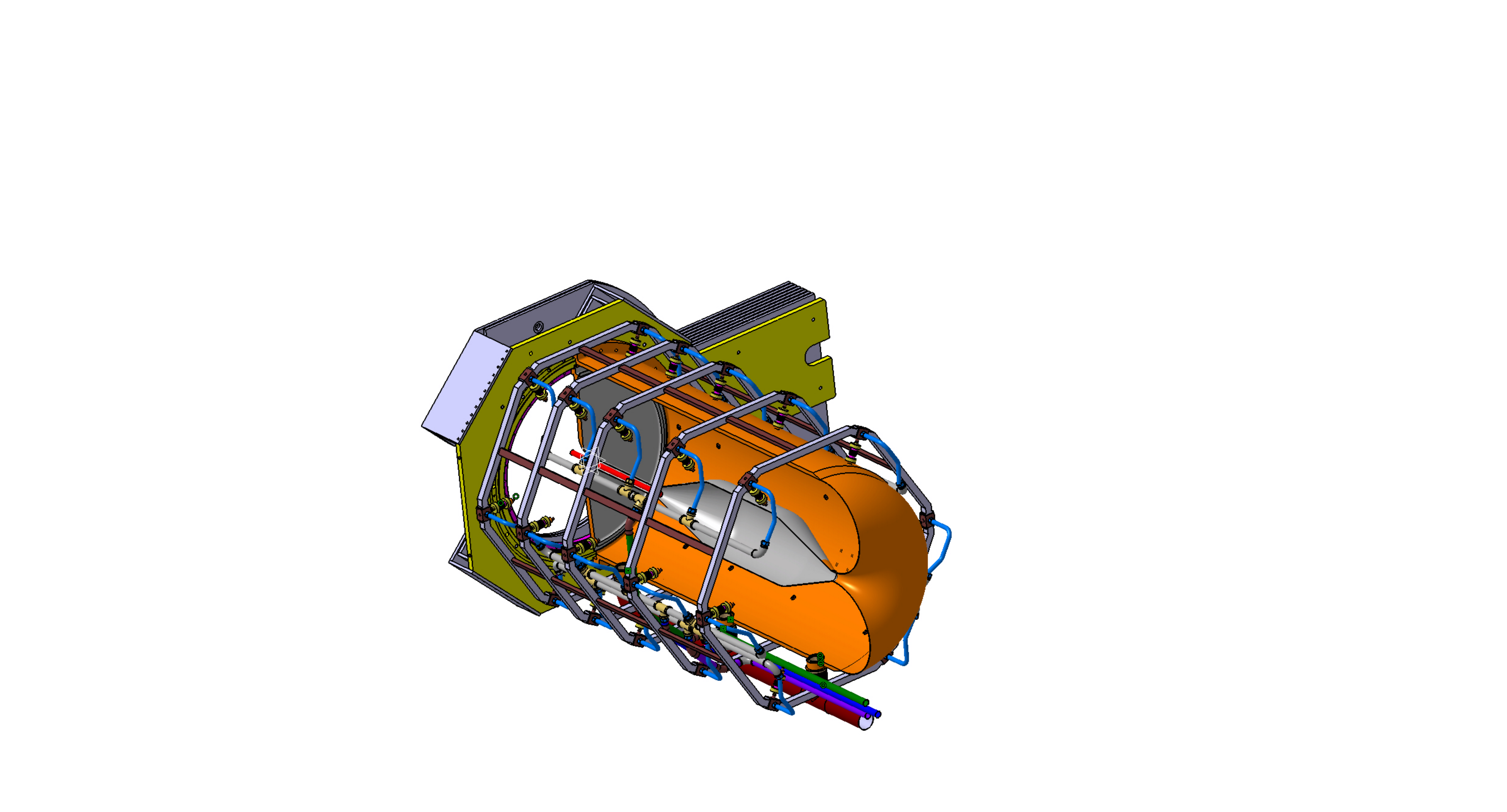}
\caption{\label{fig: horn} Horn with cooling system.}
\end{minipage} 
\end{figure}

A 4-MW proton-beam from CERN's SPL is foreseen to be separated by a series of kicker magnets into four beam lines. Then each beam will be focused by a series of quadruples and correctors to a four horn/target assembly  (figure \ref{fig: 4horn}). In this way, each horn/target assembly is able to accommodate better the multi-MW power and thus increasing its lifetime \cite{protondriver}.

A 0.25 mm thick beryllium beam window has been studied 
 as the interface between each 1 MW proton-beam line and each horn/target assembly. Maximum temperatures as high as 180 $^\circ$C and (109 $^\circ$C) and Von Mises stresses as high as 50 MPa and (39 MPa) are developed respectively for water and helium cooling: these are well below the beryllium strength limit.     
  
\section{Target studies}
A packed-bed target with Ti6Al4V-spheres and helium transverse cooling has been chosen  as the baseline target option \cite{designreport, densham}. It is placed inside the upstream part of horn's inner conductor. The advantages of the packed-bed target are among others a$)$ large surface area for heat transfer with coolant able to access areas with highest energy deposition b$)$ minimal thermo-mechanical and inertial stresses and c$)$ potential heat removal rates at the hundreds kilowatt level with high helium flow rate. Advantages of the helium transverse cooling are a$)$ almost beam neutral  b$)$ no generation of stress wave in coolant and c$)$ low activation of coolant with no corrosion problems. 

Alternatively, a pencil-like geometry of solid beryllium has been studied \cite{designreport}. This pencil-like geometry gives steady-state thermal stress within acceptable range for Beryllium. Pressurized helium cooling appears feasible but center proton-beam effects could be problematic because of the stress induced: this point needs further thermo-mechanical studies. 

\section{Horn studies}

\subsection{Horn shape and layout-geometry optimization \label{horngeometry} }

The end-design consists of an inner conductor with a cylindrically shaped upstream part to decrease the transverse momentum of the low-energy charged mesons, followed by a trapezoidal shaped middle part to select a special particle energy spectrum  (for optimum physics) and finally a convex downstream plate to de-focus wrong-sign mesons that contribute to the background neutrino spectra. 
 This configuration has been selected as the best compromise between physics performance and reliability under 1 to 1.3 MW proton-beam power \cite{designreport}. 
The detailed drawing for a horn is shown in figure \ref{fig: horn}.

The horn parameters as well as the geometrical parameters of the decay tunnel (length and radii) are optimized for the best achievable sensitivity limit on sin$^{2}2\theta_{13}$. The beam parameters are initially scanned broadly and then restrictly in three iterations in order to minimize the CP-violation averaged 99\% C.L. sensitivity limit on  sin$^{2}2\theta_{13}$ \cite{physicsopt, hornopt}. 

\subsection{Horn thermo-mechanical and dynamical stress studies}

\begin{figure}[h]
\begin{minipage}{17pc}
\includegraphics[width=17pc]{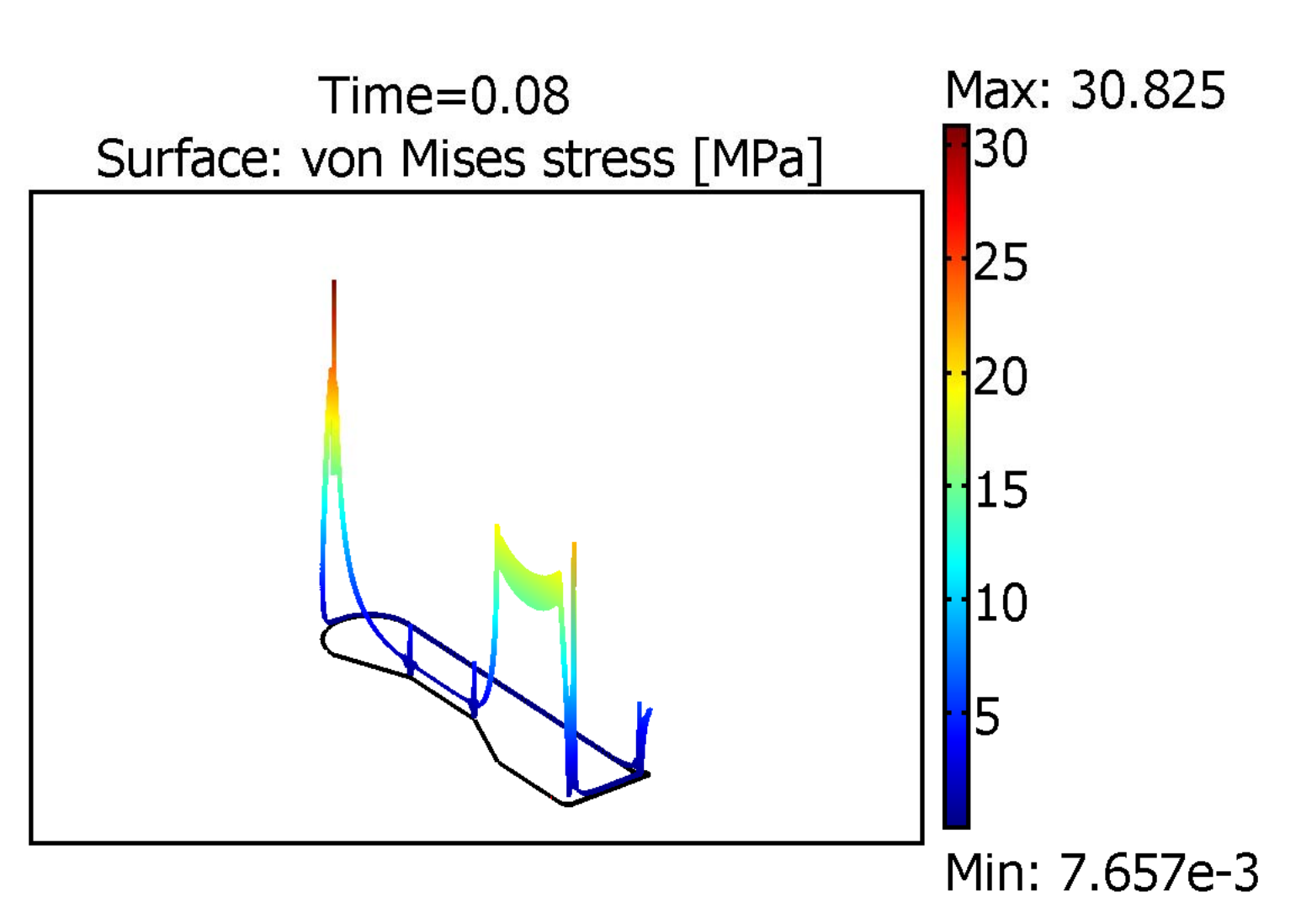}
\caption{\label{fig: h_stress} s$_{max}$=30 MPa.}
\end{minipage}\hspace{2pc}%
\begin{minipage}{18pc}
\includegraphics[width=18pc]{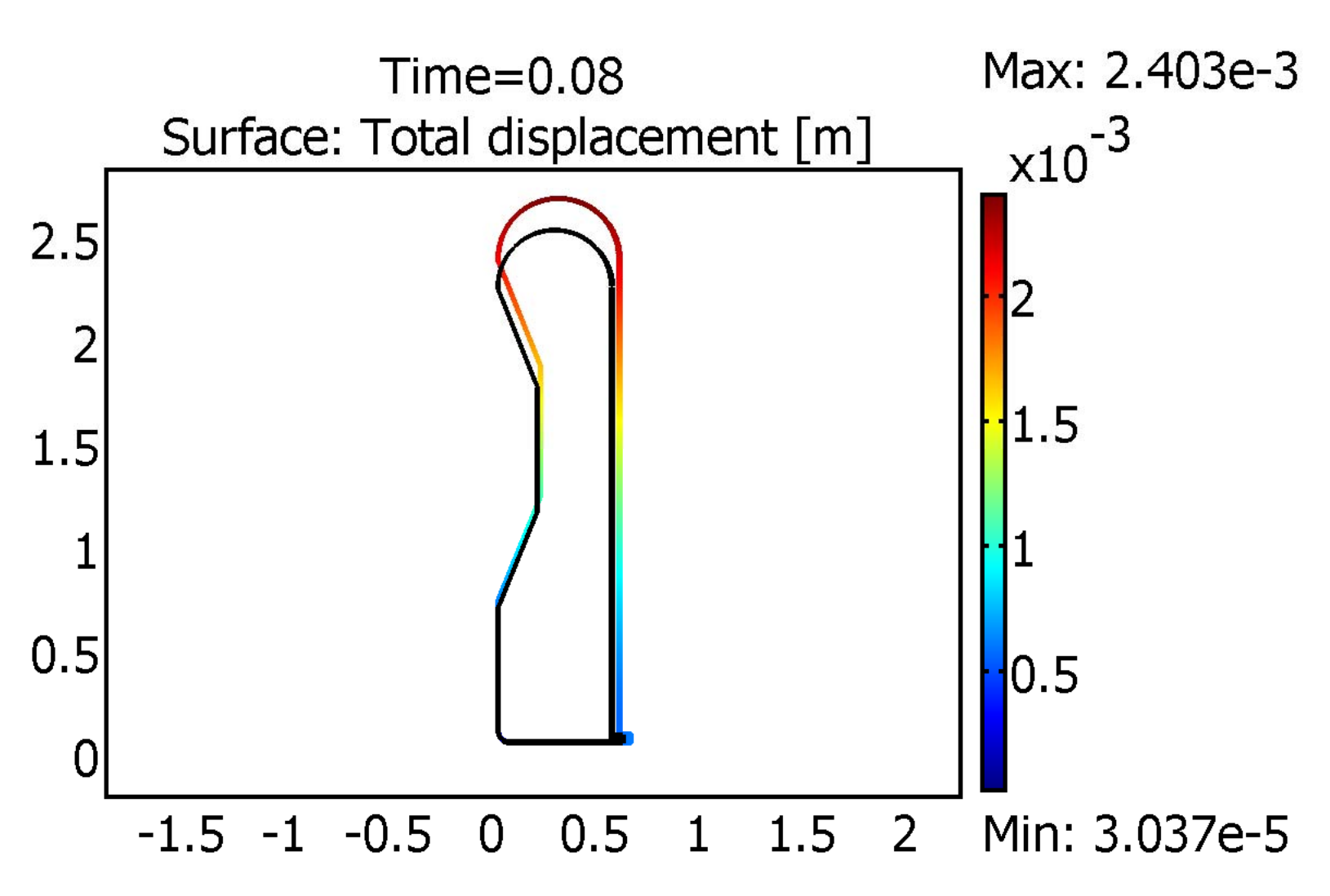}
\caption{\label{fig: h_dis} u$_{max}$=2.4 mm.}
\end{minipage} 
\end{figure}

The Aluminum alloy Al-6061T6 horn is subjected to cyclic deformation due to a pulsed magnetic pressure load. In addition, the temperature field creates a thermal static stress. For a non uniform cooling with maximal temperature of 60 $^\circ$C, the maximal static thermal stress is calculated about $60$ MPa and is located in the upstream corner and downstream top part of the horn. If a uniform temperature is achieved everywhere, the horn is expanding, and the maximum thermal static stress is 6 MPa \cite{hornstudies}. The combined stresses in the inner conductor due to the magnetic pulses and the thermal stress for a uniform achieved temperature of $60$ $^\circ$C are around $20$ MPa. For the inner conductor horn, the magnetic pressure pulse creates a peak Von Mises stress of about $16$ MPa. The stress and the deformation are shown in figures  \ref{fig: h_stress} and \ref{fig: h_dis}.

\subsection{Cooling system}
To remove a total power of about $60$ kW and maintain a temperature of about $60\,^\circ $ C a water-jet cooling system is being studied.  This system will be made of 5 rows of 6 nozzles (figure \ref{fig: horn}) and to spray water toward the inner conductor of the horn. The estimated water flow rate is calculated between 60 to 120 l/min per horn depending the design.
To minimize the thermal static stress, the nozzle size and disposition should be properly located in order to achieve the most uniform temperature everywhere inside the horn \cite{cooling}. 


\subsection{Fatigue}
There is no fatigue limit available for Aluminum alloy, which means that fatigue data can only give a probability of failure for a determined level of stress and number of cycles. 
 According to \cite{fatigue_curves}, the fatigue strength limit is 20 MPa for 10$^9$ pulses with a mean static stress due to thermal dilatation. For the weld junction a limit of $10$ MPa should be respected to maximize horn lifetime \cite{hornstudies}.

\subsection{Power supply}
A one-half sinusoid current waveform with a 350 kA maximum current and pulse-length of 100 $\mu$s at 12.5 Hz frequency is needed for each horn. In order to compensate for these extreme conditions a  power supply with 10 modules branched in parallel with 35 kA current each running at 50 Hz frequency is being studied. Also, the power supply components lifetime is very important as 
0.216 10$^{9}$ pulses for each horn's switch and a total of  0.864 10$^{9}$ pulses per 200 days for the power supply are expected.  
Thus, the following designs are being investigated: a power supply  with energy recovery made by a) a self that implies huge stresses on the switches and the capacitor due to high reverse voltage and current and  b) a diode that implies less stress on the switches and the capacitor, but increase the electric consumption and energy thermal dissipation. Their comparison in terms of manufacturing, reliability and electric costs will define the final solution.

\section{Safety}



The future design of Multi-Mega Watt sources facility has to take to account the significant amount of radiation produced during beam operation and the radio-activation of the surrounding environment. The design of the shielding should reduce the dose equivalent rate to a minimal level. In order to reach these dosimetry objectives the ALARA (As Low As Reasonably Achievable) approach will be used in the design of the facility. ALARA consists of an iterative process between three phases: a) Preparation,  design of the facility,  dose equivalent rate map,  study the intervention  procedures for workers b) execution, engineering phase check/improve the dosimetry objectives and c) analyse and feedback on safety from others experiments.

\begin{figure}[h]
\begin{minipage}{16pc}
\includegraphics[width=14pc]{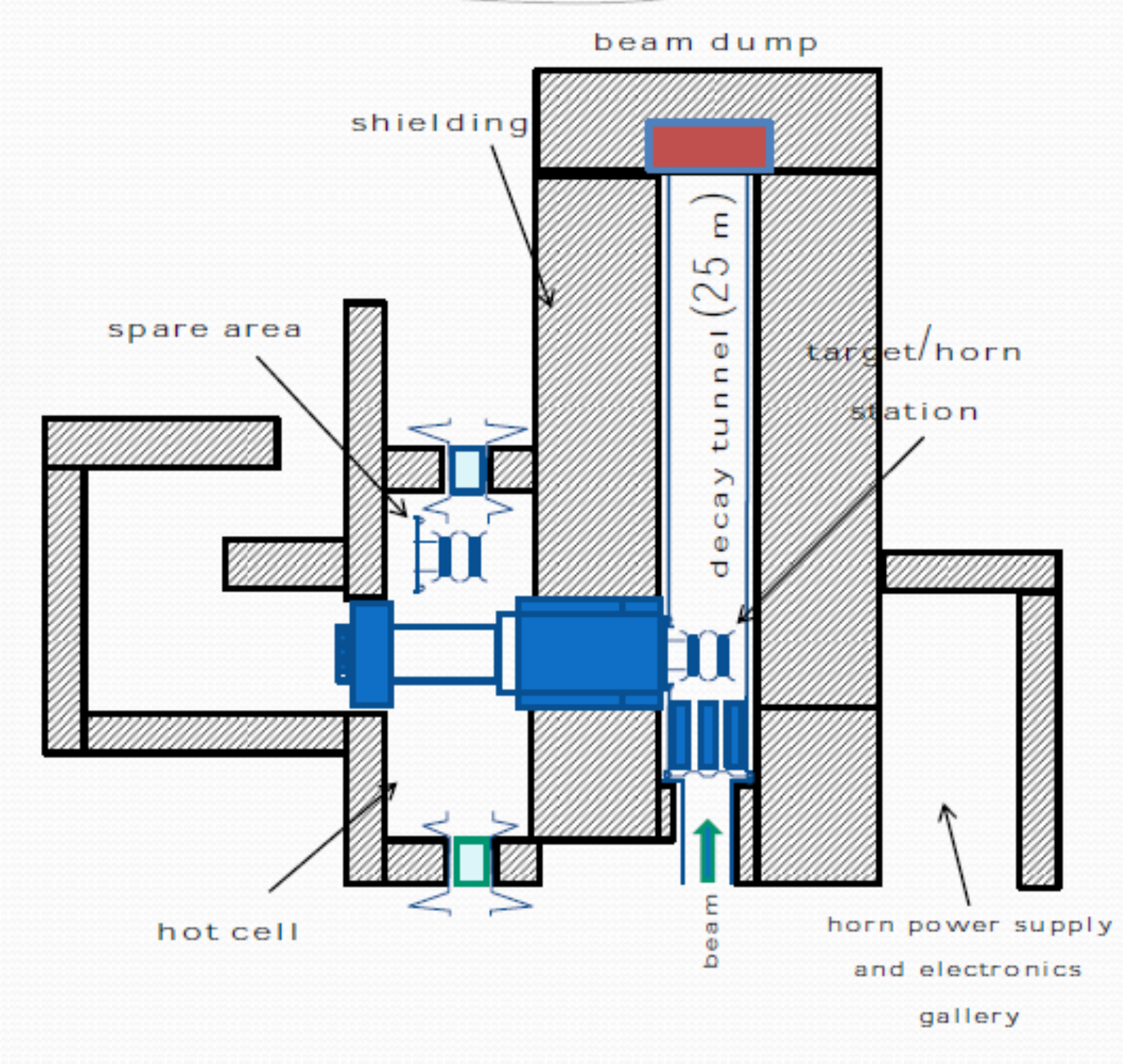}
\caption{\label{fig: newlayout} Super Beam infrastructure.}
\end{minipage} \hspace{2pc}%
\begin{minipage}{19pc}
\includegraphics[width=19pc]{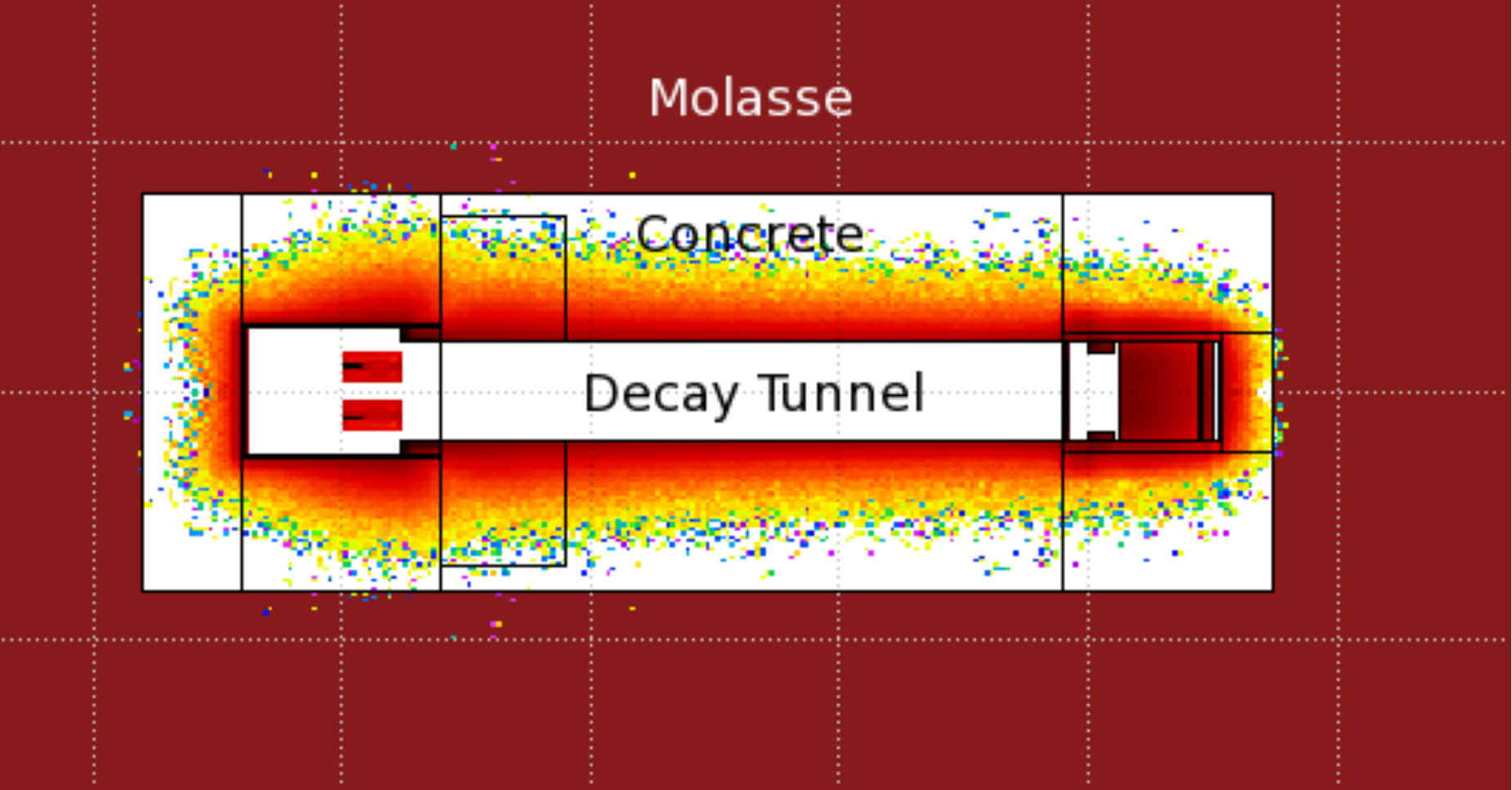}
\caption{\label{fig: activity} Activity distribution in facility.}
\end{minipage}
\end{figure}

The Super Beam infrastructure consists of: a) Proton Driver line, b) Experimental Hall (Target Station, Decay Tunnel, Beam Dump), c) Spare Area Room, d) Hot Cell, e) Service Galleries (Power supply, Cooling system) f) Air-Ventilation system and g) Waste Area (figure \ref{fig: newlayout}). 

As example of safety Monte-Carlo studies with FLUKA, the activation of the whole structure and of molasse rock surroundings have been evaluated after 200 days of irradiation: there is minimum rock (molasse) activation surrounding the Super Beam facility due to concrete shielding with 5.5 m of thickness as shown in figure \ref{fig: activity}.  

\section{Conclusion}

Monte-Carlo, thermo-mechanical and dynamical stress finite-elements analysis studies show that the four-horn/target system can be feasible under the extreme 4 MW proton-beam power conditions. Furthermore, R\&D is needed for the target and the horn in order to study the fatigue, cooling, power supply designs and radiation degradation.    

\ack We acknowledge the financial support of the European Community under the European Commission Framework Programme 7 Design Study: EUROnu, Project Number 212372. The EC is not liable for any use that may be made of the information contained herein.


\end{document}